\documentclass[pre,twocolumn,superscriptaddress,showpacs]{revtex4}

\usepackage{graphicx}
\usepackage{amsmath}

\newcommand{\rmd}{\ensuremath{\mathrm{d}}}

\newcommand{\G}{\mathcal{G}}

\begin{document}

\title{Free energy landscapes, dynamics and the edge of chaos in mean-field 
models of spin glasses }
\date{\today}

\author{T.\ Aspelmeier}
\affiliation{Institut f\"ur Theoretische Physik,
Universit\"at G\"ottingen, Friedrich-Hund-Platz\ 1, 37077 G\"ottingen, Germany}

\author{R.\ A.\ Blythe}
\affiliation{School of Physics and Astronomy, University of 
Manchester, Manchester M13 9PL, U.K.}
\affiliation{School of Physics, University of Edinburgh, Mayfield
  Road, Edinburgh EH9 3JZ, U.K.}

\author{A.\ J.\ Bray}
\author{M.\ A.\ Moore}
\affiliation{School of Physics and Astronomy, University of 
Manchester, Manchester M13 9PL, U.K.}
 
\begin{abstract}
Metastable states in Ising spin-glass models are investigated
numerically by finding iterative solutions of mean-field equations for
the local magnetizations $m_i$. A number of iterative schemes are
employed, and two different mean-field equations are studied: the
Thouless-Anderson-Palmer (TAP) equations that are exact for the
Sherrington-Kirkpatrick model, and the simpler `naive-mean-field'
(NMF) equations, in which the Onsager reaction term of the TAP
equations is omitted and which are exact for the Wallace model.  The
free-energy landscapes that emerge are  very different for the two
systems.  For the TAP equations, the numerical studies confirm the
analytical results of Aspelmeier et al., which predict that TAP states
consist of close pairs of minima and index-one (one unstable
direction) saddle points, while for the NMF equations the
corresponding free-energy landscape contains saddle points with large
numbers of unstable directions. For the TAP equations the free energy
difference between a minimum and its adjacent saddle point (the
`barrier height') scales as $1/(f-f_0)^{\frac{1}{3}}$ where $f$ is the
free energy per spin of the solution and $f_0$ is the equilibrium free
energy per spin.  This means that for `pure states' for which $f-f_0$
is of order $1/N$, where $N$ is the number of spins in the system, the
barriers between them scale as $N^{\frac{1}{3}}$, but between states
for which $f-f_0$ is of order one, then the barriers are finite and
also small so such metastable states will be of limited physical
significance. For the NMF equations there are saddles of index $K$ and
we can demonstrate that their complexity $\Sigma_K$ scales as a
function of $K/N$.

We have also employed an iterative technique with a free parameter
that can be adjusted to bring the system of equations close to the
`edge of chaos'.  Both for the TAP and NME equations it is possible
with this approach to find metastable states whose free energy per
spin is close to $f_0$.  As $N$, the number of spins is increased, it
becomes harder and harder to find solutions near to the edge of chaos,
but nevertheless the results which can be obtained are competitive
with those achieved by more time-consuming computing methods and
suggest that this method may be of general utility.

\end{abstract}

\pacs{75.10.Nr, 75.50.Lk, 05.70.Ln}

\maketitle

\section{Introduction}

The study of metastable states in spin glasses began some 25 years ago
with the calculations of Tanaka and Edwards \cite{TE} (TE) and of Bray
and Moore \cite{BM} (BM), dealing with metastable states at zero and
non-zero temperatures respectively, within the infinite-range
Sherrington Kirkpatrick model.  For the zero-temperature studies,
metastable states were defined as states which are one-spin-flip
stable, i.e.\ states for which flipping any one spin increases the
energy, while for $T>0$ they were identified with solutions of the TAP
equations \cite{TAP}, which are exact for the SK model.  While the TE
calculation is quite straightforward, the original BM calculation
involved some technical challenges which were finessed in the first
attempt, and have only recently been satisfactorily resolved, some 24
years after the original paper \cite{ABM04,CGP04,Rome}.  While the
central result of BM, that the mean number (averaged over disorder
configurations) of TAP solutions increases exponentially with the
number of spins, $\langle N_s \rangle \sim \exp[N\Sigma(T)]$, where
$\Sigma(T)$ is the `complexity', was confirmed by the later work,
new insights into the nature of the TAP solutions and the structure of
the free-energy landscape were obtained.  Specifically, it was found
that all solutions correspond either to minima of the TAP free energy
$F_{\rm TAP}$, or to saddle points of index one, where the index of a
saddle point is the number of negative eigenvalues of the Hessian
matrix $\partial^2F_{\rm TAP}/\partial m_i\partial m_j$, so that
minima have index zero.  Furthermore, the minima and saddle points
occur in close (in configuration space) pairs \cite{ABM04}.  This
structure has implications for the dynamics in the SK model. In
particular we will study the free energy difference between the saddle
and its associated nearby minimum.  This difference would seem
intuitively to be related to the barrier which has to be overcome to
escape from a pure state and we will find that it increases with $N$
like $N^{1/3}$. However, for the vast majority of the TAP states---all
those with free energy per spin larger than its equilibrium
value---the barriers are finite. This means that such TAP states are
probably of little physical significance for the dynamics of the SK
model.

We will contrast the TAP free-energy landscape with that of the naive
mean-field (NMF) equations. We show numerically that the free-energy
function $F_{\rm NMF}$ possesses saddle points with large indices $K$
(up to a maximum proportional to $N$), and we show that the
corresponding landscape is more rugged than the TAP landscape and
tends to trap the `iterative' dynamics at (or close to) a threshold
free energy where the minima numerically dominate the saddle points,
much as in the $p$-spin spherical model \cite{pspin}.
  
This qualitative distinction between dynamics on the TAP and NMF free
energy surfaces is interesting because both sets of equations become
{\em the same} at $T=0$. In particular, they have the same ground
states. This suggests a program for finding the ground-state energy by
working with the TAP equations at very low but non-zero temperature,
and using the resulting states at starting configurations for a final
quench at $T=0$. In view of the preceding discussion, the same
approach applied to the NMF equations would not be successful, since
the system would get stuck at a threshold free energy, as for the
$p$-spin spherical model.  If one hopes to determine ground-state
properties using $T>0$ equations, therefore, it is important to choose
equations with a favorable free-energy surface (like TAP) rather than
an unfavorable one (like NMF). In other words if one is trying to find
low energy states of (say) the SK model using standard algorithms such
as gradient descent on a free energy surface, then one gets much lower
energy solutions if one uses the exact free-energy surface than when
one uses only an approximation to it. We suspect this point will have
validity beyond the SK model.

However, we have discovered that it is possible to obtain solutions of
low free energy by a method which takes one entirely off the free
energy surface.  These states are accessed through a novel iterative
algorithm containing one adjustable parameter. We show that there is a
critical value of this parameter separating runs which terminate from
those which do not, and that the lowest free-energy states are
accessed when the parameter is close to the critical value.  As the
parameter is increased towards the critical value, one sees many
solutions which are limit cycles of a length which also increases as
the critical value is approached and beyond the critical value only
`chaotic solutions' seem to exist.  In the course of the iteration
it is possible for the magnetizations to take unphysical values such
that $|m_i|>1$.  We speculate that this approach may be a generic
technique for finding low-cost states in hard optimization
problems. Somehow being close to the `edge of chaos' enables one to
explore phase space and it is only the lowest states, which are
surrounded by the highest barriers, which can trap the algorithm into
a fixed point rather than a limit cycle.  Possibly related ideas are
in a paper by Boettcher and Frank \cite{BF}.

The outline of the paper is as follows.  In section \ref{section2} we
state the TAP and NMF equations, discuss the algorithmic methods we
will adopt and demonstrate that for NMF most of the standard
algorithms seem to terminate at a free energy which is relatively
insensitive to the algorithm.  In section \ref{exhaustive}, we explain
this result by carrying out an exhaustive search for turning points of
the free energy for small systems, showing that there is a threshold
free energy below which minima numerically dominate saddle points.
The connection between the structure of the free-energy landscape and
the nature of the dynamical attractors is further explored in section
\ref{connection}. Section \ref{sec:TAP} contains a detailed study of
low free-energy TAP states obtained from studying the solutions using
the `edge of chaos' algorithm. Section \ref{discussion} concludes
with a discussion and summary of the main results.

\section{Solving the TAP and NMF equations by iteration}
\label{section2}

The  TAP  and  NMF  equations   can  be  derived  from  the  following
generalized free-energy function:
\begin{multline}
\label{freeenergy}
F  = - \sum_{(i,j)} J_{ij}  m_i m_j  - \gamma  \frac{\beta N}{4}
(1-q)^2  + {}  \\  \frac{1}{\beta} \sum_i  \left[ \frac{1+m_i}{2}  \ln
\frac{1+m_i}{2} + \frac{1-m_i}{2} \ln \frac{1-m_i}{2} \right]
\end{multline}
where $\gamma=1$  corresponds to TAP  and $\gamma=0$ to NFM.   In Eq.\
(\ref{freeenergy}) the local magnetizations, $m_i = \langle S_i \rangle$,
lie in  the range $-1  \le m_i  \le 1$, $\beta  = 1/kT$ as  usual, and
$q=(1/N)\sum_i m_i^2$.

The TAP  and NFM equations  themselves are derived from  the equations
$\partial{F}/\partial m_i = 0$, $i=1,\ldots,N$, which give the turning
points of $F$. They can be written in the form
\begin{equation}
\label{turningpoint}
m_i =  \G_i(\vec{m}) = \tanh \left[  \beta h_i -  \gamma \beta^2 (1-q)
m_i \right]
\end{equation}
where the  local field $h_i =  \sum_{j \ne i} J_{ij}  m_j$. The second
term in  the argument  of the  $\tanh$ function is,  for the  TAP case
$\gamma=1$, the Onsager reaction term. The TAP equations are exact as
$N \rightarrow \infty$ for the SK model. The naive mean-field equations
are exact for the Wallace model, in which at each site $i$ in the system there
 is a set of Ising spins $s_{ia}$ ($a=1,2, \ldots,k)$, each of which
interacts with the $k$ spins at sites which are coupled to the $i$-th site and
the limit $k\rightarrow \infty$ is taken \cite{BSY86}.

The set of spins $m_i$ that satisfy (\ref{turningpoint}) can be
thought of as a fixed point $\vec{m}^\ast$ of the map
\begin{equation}
\label{simplemap}
m_i^{(k+1)} = \G_i(\vec{m}^{(k)}) \;.
\end{equation}
This  map could  be applied  to all  spins in  parallel, to  each spin
separately and in  sequence, each time to a random spin,  or to a spin
chosen  for some particular  reason.  It  turns out  that for  the NMF
equations  ($\gamma=0$), application  of this  map leads  always  to a
fixed  point  that  coincides  with  a  minimum  in  the  free  energy
landscape,  no  matter  how  the   map  is  applied.   This  fact  one
establishes by calculating, at the fixed point, the Hessian $H$ which 
is the matrix of second derivatives of the free energy,
\begin{multline}
\label{hessian}
H_{ij}   =   \left.    \frac{\partial^2   F}{\partial   m_i   \partial
m_j}\right|_{\vec{m}^\ast} =  - J_{ij} -  \gamma \frac{2\beta m^\ast_i
m^\ast_j}{N} + {}\\  \left[ \frac{1}{\beta} \frac{1}{1-(m^\ast_i)^2} +
\gamma \beta(1-q) \right] \delta_{ij},
\end{multline}
(which holds for all $i,j$ if one takes $J_{ii} \equiv 0$) and
checking that all its eigenvalues are positive, thus defining a
minimum.  As we will see below, the free energy landscape also has
turning points with a positive number $K$ of negative eigenvalues---we
call these saddle points of index $K$.  On the other hand the
inclusion of Onsager reaction term leads to the map (\ref{simplemap})
almost never converging on a fixed point.

Clearly one does not imagine that the trajectory described by
iteration of the map (\ref{simplemap}) corresponds to any reasonable
physical dynamics on the free energy landscape, e.g., a
zero-temperature dynamics in which one follows the path of steepest
descent to a local (in general, metastable) minimum.  Unfortunately,
numerical algorithms that one might expect to approximate closely a
physical dynamics (such as steepest descents or conjugate gradients)
turn out to be poorly adapted to the NMF free energy landscape.  The
reason for this is that the free energy minima tend to lie close to
the sides of the hypercube $-1 \le m_i \le 1$ over which the free
energy is defined.  Naively, one might imagine that the divergence in
the free energy gradient as the hypercube's boundary is approached
would prevent a descent algorithm from exiting the physical region.
However, the weak and short-ranged logarithmic divergence is not
resolved numerically.  This can cause points to be reached that lie
outside the hypercube and for which the free energy (\ref{freeenergy})
is undefined, thereby creating an obvious numerical problem.  We tried
a number of modifications here to work around this, such as mapping
the bounded hypercube to an infinite domain, or inventing a ficticious
free energy outside the hypercube, all to no avail.

For   this   reason,   we   modified   instead   the   iterative   map
(\ref{simplemap}) in a way that we believe gives some insight into the
free energy minima located by a physical dynamics.  This map reads
\begin{equation}
\label{bettermap}
m_i^{(k+1)}  =   m_i^{(k)}  +  \alpha   \left[  \G_i(\vec{m}^{(k)})  -
  m_i^{(k)} \right]
\end{equation}
in which $\alpha$ is a parameter that controls the speed of approach
to the next iterate.  We describe the utility of this more general
iterative scheme first in the context of the NMF equations
($\gamma=0$) and then the TAP equations ($\gamma=1$).

\subsection{NMF Equations}

If in Eq.~(\ref{bettermap}) the parameter $\alpha$ is small, one
expects a smooth trajectory to be followed, and crucially one that is
constrained to remain within the hypercube.  This statement can be
made more precise by noting that the displacement $\delta m_i = \alpha
(\tanh \beta h_i - m_i )$ is bounded: $|\delta m_i| \le 2 \alpha$ and
so can be made arbitrarily small by decreasing $\alpha$.  Furthermore,
one can show that the projection of this displacement onto the
gradient vector of the free energy is always negative.  Thus, for
sufficiently small $\alpha$, the dynamics should take a small step in
such a way as to lower the free energy, a fact that has been verified
numerically.  Although we do not claim that this dynamics is
equivalent in any sense to a steepest-descent dynamics, the fact that
the free energy is lowered in a controlled fashion leads us to view it
as a `physical' dynamics.

In  Table~\ref{NMFalgorithms} we display  the free  energy of  the NMF
minima located  using four different algorithms, averaged  over both a
number of  different realizations of  the disorder (i.e., sets  of the
random  variables   $J_{ij}$,  each  of   which  we  refer  to   as  a
\emph{sample}) and  over a number  of different initial  conditions on
the iteration.  So that the  algorithms can be properly compared for a
given system  size $N$, each was  treated to the same  set of samples.
The system sizes  used were quite small ($N=20,30,40$)  so that we may
later compare with the exhaustive search for \emph{all} turning points
in the  same samples (see Section~\ref{exhaustive}).   The first three
of  the  four  algorithms  are  implementations  of  (\ref{simplemap})
applied  on   a  spin-by-spin  basis   in  different  ways.    In  the
\emph{random}  algorithm, the  spin  chosen for  update  is chosen  at
random,  whereas in the  \emph{cyclic} algorithm,  each is  visited in
order.  Meanwhile,  the spin chosen in the  \emph{greedy} algorithm is
that for which the free energy will decrease the most under application 
of the map.  Finally
the   \emph{physical}   algorithm   uses   (\ref{bettermap})   applied
\emph{simultaneously}  to all  spins  and with  the control  parameter
$\alpha=10^{-3}$.  We  notice from the  Table that the  differences in
the free energy arising from a change of algorithm, whilst significant
compared to the errors, are about two orders of magnitude smaller than
those arising  from an  increase in the  system size.  Later,  we will
provide evidence  that these  changes are also  small on the  scale of
free energies spanned by all the solutions of the NMF equations.

\begin{table}
\begin{tabular}{c|c|c|c|c|c}
$N$ & Samples & Random & Cyclic & Greedy & Physical \\\hline\hline
20 & 2000 & -0.67196(3) & -0.67151(3) & -0.67285(3) & -0.67211(3)\\
30 & 1000 & -0.68811(4) & -0.68760(4) & -0.68882(4) & -0.68804(4)\\
40 & 1000 & -0.69900(4) & -0.69862(4) & -0.69973(4) & -0.69892(4)
\end{tabular}
\caption{\label{NMFalgorithms}Average free energy of minima in the NMF
  free energy landscape reached using various iterative algorithms.
  For all system sizes $N$, $\beta=2$ and different minima were
  located by restarting the algorithm for each sample from 5000
  different, random initial conditions.}
\end{table}

\subsection{TAP Equations}

Solutions  of  the  TAP  equations,  i.e.\  (\ref{turningpoint})  with
$\gamma=1$, are much  harder to find than those  of the NMF equations.
Part of the reason for this is that including the Onsager reaction
term  stabilizes  the  trivial  (paramagnetic)  solution  $\vec{m}=0$.
Indeed,  when  we  employed  simple-minded  free  energy  minimization
algorithms on  the TAP free  energy landscape, we found  almost always
this    trivial   solution.  However,   this    solution   is
\emph{thermodynamically} unstable in the spin glass phase \cite{MPV87} 
and  so one must  take steps  to avoid  it.  One  possibility, pursued
elsewhere, is  to use a different  free-energy function  in the 
thermodynamically unstable region \cite{Plefka02}.

We succeeded in finding solutions of the TAP equations using the
iterative map (\ref{bettermap}) in the regime where $\alpha$ is
\emph{greater} than one.  The reason why this choice is useful is that
with $\alpha\le1$ the iteration enters a limit cycle between points
that are unrelated to the turning points of the TAP free energy
landscape.  By amplifying the displacements between these points, it
is possible to break out of the limit cycle.  An added bonus is that
paths to the trivial solution $\vec{m}=0$ are also destabilized---this
is because our observations suggest paths to this solution have an
oscillatory character.

Clearly, there is an optimum value for $\alpha$ for finding the
largest number of solutions of the TAP equations.  Too small, and the
limit cycles are attractive; too large, and one may hop in great leaps about the hypercube, never to settle on a solution.  As we shall see in Sec.~\ref{sec:TAP},
the optimum $\alpha$ is in fact close to a point where the iterative
dynamics become chaotic.  As this point is approached, the dynamics do
exit the hypercube, but there are no numerical problems associated
with this since the right-hand side of (\ref{bettermap}) is defined
even if $|m_i|>1$.  We only require that the trajectory followed by
the iteration converges on a fixed point inside the hypercube in a
finite time.  As we shall also later explain in detail, when $\alpha$
is tuned to its optimum we can approach closely states with the
thermodynamic equilibrium free energy.

\subsection{Preliminary Comparison of TAP and NMF}

\begin{figure}
\label{QuenchedStates}
\includegraphics[width=\linewidth]{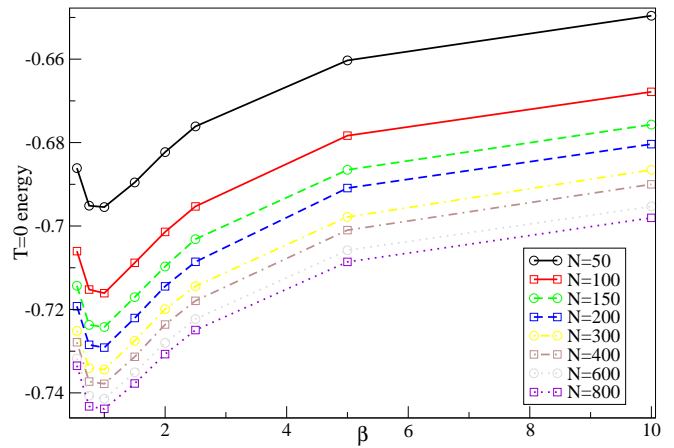}
\caption{Energies reached  from solutions of the
NMF equations found by iteration after  a quench to a $T=0$ state at
different system size $N$ and inverse temperature $\beta$.}
\end{figure}

We believe that, despite having the same $T\to0$ limit as the TAP
equations, solutions of the NMF equations will not take us to the
lowest energy states.  Unfortunately, since the equilibrium free
energy of the NMF equations is not known exactly at finite
temperature, we are unable to test whether it can be reached for this
case.  Instead, we probe the relationship between NMF free energy
minima and low-lying states by performing a zero-temperature quench
from the former.  Specifically, having found a solution to the NMF
equations, i.e., (\ref{turningpoint}) with $\gamma=0$, at finite
temperature $\beta$, we then iterate the $T=0$ version of the map.
That is, every spin is set to have unit magnitude but with the sign
taken from the final state of the NMF iteration.  Then, sites are
visited randomly and flipped if they are not aligned with their local
fields until a metastable state is reached in which all spins have
become aligned with their fields.  The energies of these states,
averaged over different NMF turning points and different bond
distributions, are shown in Fig.~\ref{QuenchedStates} for a range of
system sizes $N$ and inverse temperatures $\beta$.  A noticeable
feature of the graph is that these energies have a nonmonotonic
dependence on $\beta$ and \emph{increase} as the temperature is
decreased past a certain point.  This suggests that minima of the NMF
free energy landscape (at least, those found by iterative methods) at
low temperatures are far away from the low energy states.  This one
sees from the fact that the lowest energy states obtained by this
process occur at an apparently well-defined \emph{finite} temperature
$\beta \approx 1$.

A plausible explanation for the minimum is as follows. The total
number  of  NMF  solutions  has  the  form  $\exp[N\Sigma(T)]$,  where
$\Sigma(T)$ is  expected to be a monotonically  decreasing function of
$T$,  vanishing at the  critical temperature  $T_c=2$ (where  there is
only  one solution,  $m_i=0$ for  all  $i$). Exactly this behavior  is
obtained for the  TAP equations \cite{BM}, except that $T_c=1$ for TAP. 
It follows that, for any given system size $N$, there will be a range of
temperatures  near $T_c$ where  the condition  $N\Sigma(T) \ll  1$ will
hold.  In  this regime  there  will be  a  unique  solution with  high
probability.  In  fact  there  will  be $O(1)$  solutions  down  to  a
temperature $T^*(N)$  defined by $N\Sigma(T^*) = 1$.   For $T>T^*$ the
free energy will  be determined essentially exactly, as  there is only
one solution. This state should  provide a good starting point for the
subsequent quench to  $T=0$. Our hypothesis, then, is  that the minima
in Figure~\ref{QuenchedStates} occur at $T \simeq T^*(N)$. The relative 
insensitivity to $N$ of the position of the minimum in 
Figure~\ref{QuenchedStates} may be accounted for by the strong  
$T$-dependence of  $\Sigma(T)$ in the  regime $T  > T_c/2$
(evident in the exact solution for TAP \cite{BM}).

As we have mentioned above, solutions of the TAP equations by
(suitably chosen) iteration at the `edge of chaos' can as $T\to0$,
yield states of the SK spin glass for which the energy per spin is
close to the equilibrium value.  We have also used an `edge of chaos'
algorithm for the $T=0$ version of the NMF equations and there too
states of low free energy are obtained. However, the data presented in
Fig.~\ref{QuenchedStates} strongly suggests that this is not the case
for the NMF equations when treated by an iterative procedure which
parallels gradient descents.  We believe that this is a consequence of
the shape of the free energy landscape, which we shall now
discuss. (For the TAP equations gradient descents algorithms usually
fail to give any solution other than the trivial solution with all
$m_i=0$).

\section{The shape of the free energy landscape}
\label{exhaustive}

In this section we investigate  numerically some properties of the NMF
free-energy landscape in order to contrast with a similar study of the
TAP landscape presented in  \cite{CGP04}.  The strategy is to consider
small  systems and  locate (as  best one  can) \emph{all}  the turning
points in the landscape.  We  begin with a discussion of the numerical
procedure.

\subsection{Numerical method}

In \cite{CGP04}, turning points of the TAP free energy landscape were
found using Broyden's method \cite{Broyden65} for solving the system
of nonlinear equations (\ref{turningpoint}) with $\gamma=1$.  We found
this algorithm also to be applicable to the case $\gamma=0$, albeit in
a different implementation \cite{GSL}.  Although more sophisticated
and efficient root-finding algorithms exist, we found that they tended
to fail for much the same reasons as did steepest descent and
conjugate gradient methods for locating minima (as discussed above).

The  numerical algorithm  approaches a  single  root deterministically
from  an initial condition,  so in  order to  find different  roots, a
sequence  of random initial  conditions were  tried.  One  hopes that,
after a sufficiently large number of initial conditions, one will have
located each of the turning points in the NMF landscape at least once.
One test that reveals if any solutions are certainly outstanding is to
calculate the Morse sum
\begin{equation}
S_M = \sum_{i=1}^{N_{\rm solns}} (-1)^{K_i}
\end{equation}
where $K_i$ is the saddle index (number of negative eigenvalues of the
Hessian)   of   the   $i^{\rm   th}$   turning   point.    Topological
considerations imply that  $S_M = 1$ when the sum  is over all turning
points of the NMF free energy.  Hence, if one finds $S_M \ne 1$, there
remain solutions to be found.  The converse, however, is not true, and
so  one must  use a  little  trial-and-error to  estimate the  typical
number of initial conditions after which no new solutions are found.

Computational  constraints imply a  trade-off between  temperature and
system size.  The larger the system  size, the slower the search for a
single solution and if one works  at too high a temperature, this fact
places too  low a limit  on the typical  number of turning  points per
sample that can  be located in a reasonable  time.  Conversely, at too
low a  temperature, the number of  solutions per sample  grows so fast
with system size that one is unable to survey effectively a reasonable
range of system  sizes.  We found an agreeable  compromise occurred at
$\beta=2$ (the  critical point in NMF is  $\beta_c=1/2$) which allowed
the study  of samples up  to size $N=40$.   At this system  size there
were typically $600$ turning points  per sample, compared with $45$ at
$N=20$.   Whilst the  criterion $S_M=1$  was satisfied  for all  but a
handful of $2000$ samples with $N=20$  sites, it held only for about a
third  of the $1000$  samples at  $N=40$.  The  data presented  in the
following  are  obtained   by  amalgamating  results  from  \emph{all}
samples, including those whose solution sets were certainly incomplete
by virtue of  $S_M \ne 1$.  Excluding these  samples from the analysis
changes our results very little.

\subsection{Results}

We  first  examine  what  classes   of  turning  point  arise  in  the
free-energy landscape.   In the TAP  free energy landscape,  one finds
only minima  and index  $K=1$ saddles which  occur in pairs  (so that,
along  with  the  trivial   solution,  the  Morse  sum  is  satisfied)
\cite{ABM04,CGP04}. From  an analysis of all  the minimum-saddle pairs
for small systems, Cavagna \textit{et al.}\ \cite{CGP04} find that the
free energy difference between  a minimum and its corresponding saddle
point seems to vanish in the thermodynamic limit.

\begin{figure}
\includegraphics[width=\linewidth]{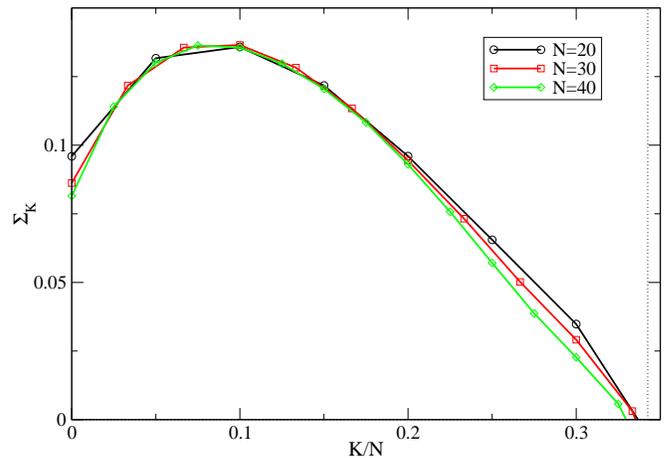}
\caption{\label{ComplexityPerIndex}Complexity  $\Sigma_K$  of the  NMF
  equations  with solutions separated  according to  the index  of the
  saddle $K$.  Near the peak, the  data collapse when $K$ is scaled by
  $N$ for  $N=20,30,40$ at $\beta=2$.  The dotted  line corresponds to
  the index of the trivial solution in the thermodynamic limit.}
\end{figure}

For the NMF equations we find minima and saddles of \emph{all} indices
up to some  maximum that grows linearly with  system size. This result
is  best illustrated  using the  \emph{saddle  complexity} $\Sigma_K$,
defined as
\begin{equation}
\Sigma_K = \frac{\ln N_{\rm solns}(K)}{N}
\end{equation}
in  which $N_{\rm solns}(K)$  is the  number of  solutions of  the NMF
equations that correspond  to a saddle point of  index $K$.  Rescaling
$K$ with  $N$ one obtains Fig.~\ref{ComplexityPerIndex}  which shows a
good collapse around the peak that occurs at $K \approx 0.1 N$.

\begin{figure}
\includegraphics[width=\linewidth]{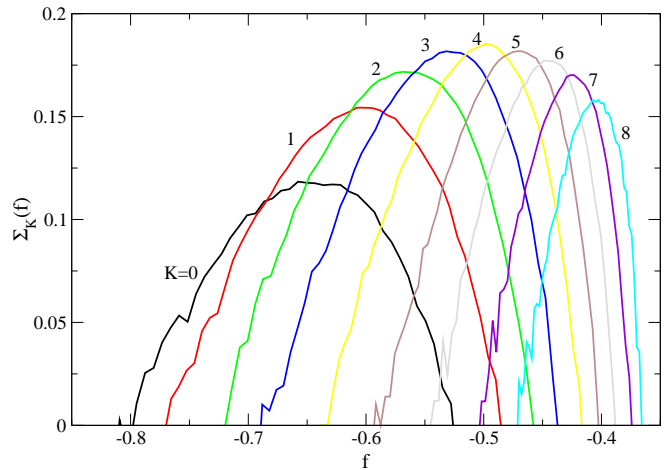}
\caption{\label{MultiComplexity}Complexity  $\Sigma_K(f)$  of the  NMF
  equations  as a  function of  free energy  $f$ for  different saddle
  index $K$, system size $N=40$ and $\beta=2$.}
\end{figure}

A little  more insight  is gained  into the shape  of the  free energy
landscape by plotting for a given saddle index $K$ the complexity as a
function of free  energy---see Fig.~\ref{MultiComplexity} for the case
$N=40$.  Here one sees a set of roughly similarly-shaped curves with a
well-defined maximum at  a free energy that increases  with $K$ whilst
the  width   of  the  curves   decreases.   This  suggests   that  the
highest-energy turning points are also the most unstable, and a closer
inspection of the data reveals  that the trivial solution seems always
to have the greatest index.

To  determine  the value  of  $K/N$ of  the  trivial  solution in  the
thermodynamic limit,  we need  to find the  total density  of negative
eigenvalues of the Hessian  matrix $H$, which from (\ref{hessian}) is,
at $\vec{m}=0$ (and $\gamma=0$ for NMF),
\begin{equation}
H_{ij} = - J_{ij} + \frac{1}{\beta} \delta_{ij} \;.
\end{equation}
In this expression, $J$ is a symmetric Gaussian random matrix, and one
can appeal  to the Wigner semicircle  law to learn that,  in the limit
$N\to\infty$, its eigenvalues $\lambda$ have a distribution
\begin{equation}
\rho_J(\lambda) = \frac{1}{2\pi} \sqrt{4 - \lambda^2}
\end{equation}
on the support  $-2 \le \lambda \le 2$. Clearly,  the diagonal part of
$H$ serves simply to shift  the distribution to the right by $1/\beta$
and so the number $K$ of the $N$ eigenvalues that are negative has the
limit
\begin{eqnarray}
\lim_{N\to\infty}  \frac{K}{N} &=&  \frac{1}{2\pi} \int_{-2+1/\beta}^0
\!\!\!  \rmd\lambda\,   \rho_J(\lambda-1/\beta)\\  &=&  \frac{1}{2}  -
\frac{1}{\pi}   \arcsin   \frac{1}{2\beta}   -  \frac{1}{4}\sqrt{4   -
\frac{1}{\beta^2}} \;.
\end{eqnarray}
For  $\beta<1/2$ we  see that  the  trivial solution  has no  negative
eigenvalues, i.e., we are above  the critical temperature for the spin
glass phase.  For the case $\beta=2$ studied numerically, the fraction
of  negative eigenvalues  approaches  $K/N \approx  0.3425$, which  is
plotted  as a vertical  dotted line  in Fig.~\ref{ComplexityPerIndex}.
We  note that  even  for  the small  systems  studied, the  complexity
approaches zero at this value.  (Actually, in these finite systems, we
obtained  solutions  with  larger   index  but  their  complexity  was
negative---i.e. there was on average  fewer than one such solution per
sample---and hence irrelevant in the thermodynamic limit).

\begin{figure}
\includegraphics[width=\linewidth]{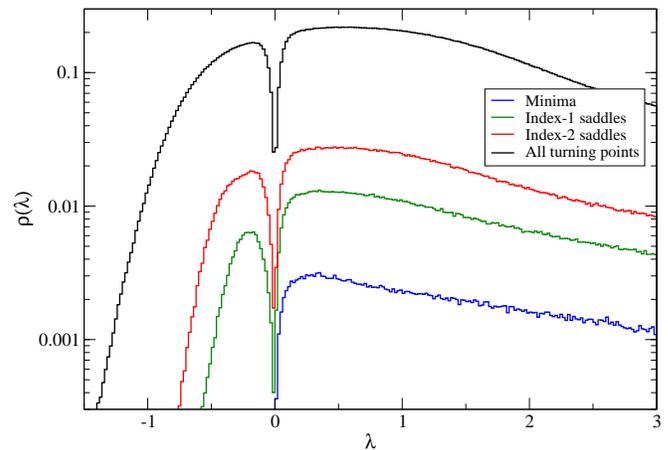}
\caption{\label{Eigenvalues}Eigenvalue distributions of the Hessian at
  minima, saddles  of index $1$ and  $2$ and all  turning points.  The
  system size $N=40$  and $\beta=2$.  Note that the  vertical scale is
  logarithmic.}
\end{figure}

More generally, one can look at the distribution of eigenvalues of the
Hessian at other turning  points.  In Fig.~\ref{Eigenvalues} these are
shown for the $N=40$ system, averaged over all turning points and also
separated out into the average  distributions at minima and saddles of
index  $K=1$ and  $2$.  Most  noticeable is  a dip  in  \emph{all} the
spectra around $\lambda=0$.  We believe  this is a real effect and not
an  artifact of  the numerics:  the  plots for  $N=20$---where we  are
confident that  all the  turning points in  the free  energy landscape
have  been  isolated---show  the   same  behavior.   We  are  however,
currently unable to explain  this repulsion of low-eigenvalue modes of
the Hessian.

\section{Connection between iterative solutions and the structure of
  the free-energy landscape}
\label{connection}

\begin{figure}
\includegraphics[width=\linewidth]{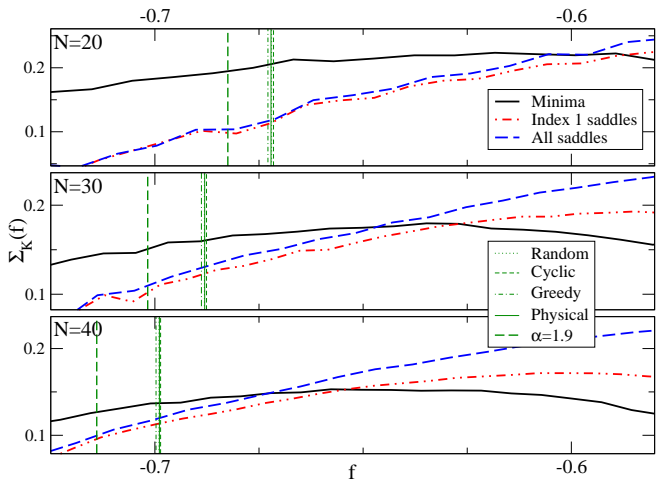}
\caption{\label{CrossingPoints}Complexity  $\Sigma_K(f)$  of  the  NMF
  equations  as a function  of free  energy $f$  for three  classes of
  turning  point (minima, $K=1$  saddles and  all saddles)  and system
  sizes $N=20,30,40$, all at $\beta=2$.  For comparison, the mean free
  energies  of states found  by the  various iterative  algorithms are
  plotted. All the energies cluster together except for those obtained
by the `edge of chaos' algorithm i.e. Eq. (\ref{bettermap}) with 
$\alpha=1.9$, which outperforms them by a considerable margin.}
\end{figure}

We  now move  on  to discuss  how  the structure  of  the free  energy
landscape influences (if at all)  the attractors of the iterative maps
described in the  previous section.  In the foregoing,  we showed that
for the NMF equations, the  typical free energies of turning points of
a particular type increase with  the saddle index $K$.  In particular,
there  is a  free energy  at which,  when approached  from  above, the
number of minima exceed the number of saddle points.  Recall, in fact,
that the  complexity describes the \emph{exponential}  increase of the
number   of  solutions   with  system   size,  so   that   below  this
characteristic free energy the ratio of the number of saddle points to
minima     vanishes     in     the    thermodynamic     limit.      In
Fig.~\ref{CrossingPoints} we  plot the mean  free energies set  out in
Table~\ref{NMFalgorithms}   for  the   various   iterative  algorithms
alongside  the corresponding complexity  histograms for  minima, $K=1$
saddles and all saddles for the three system sizes $N=20,30,40$.

\begin{figure}
\includegraphics[width=\linewidth]{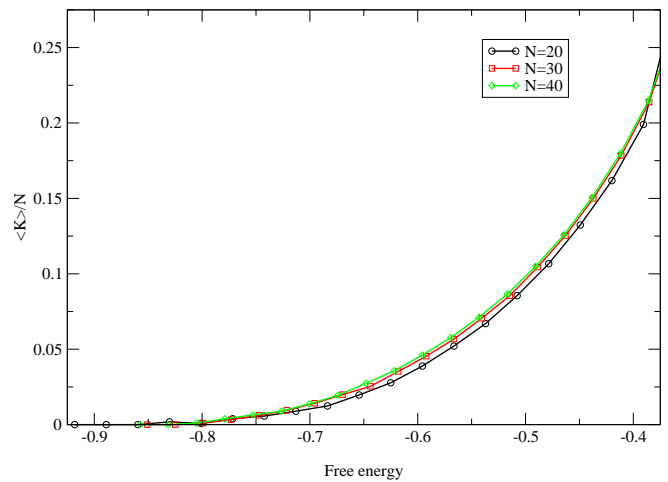}
\caption{\label{indexdensity}Index density   of  the  NMF
  equations  as a function  of free  energy $f$ at $\beta=2$  and system
  sizes $N=20,30,40$. }
\end{figure}

We first notice that the spread in free energies within the different
iterative algorithms (indicating a slight bias of the algorithms
towards different attractors) is small compared to the free-energy
range spanned by a set of turning points (the minima, say).  Noting
that all three plots in Fig.~\ref{CrossingPoints} have the same
horizontal scale, it is clear that as $N$ increases, the mean free
energy located by the iterative algorithms approaches the largest free
energy at which the minima start to outnumber the saddles.  Of course,
since we have been unable to access larger system sizes this provides
no proof that the these points will coincide as $N\to\infty$; however,
this data provides support for a `landscape' picture of
zero-temperature dynamics in which the free energy is lowered until
such time as the minima dominate to such an extent that becoming
trapped in a local minimum is an inevitability. Notice that in the
`saddles rule' type of hypothesis \cite{GCGP02} the dynamics is
supposed to be trapped at the value of the free energy at which the
index density $k(f)=\langle K\rangle/N$ becomes non-zero, where the
average is over all solutions whose free energy lies between $f$ and
$f+df$. We have plotted this quantity in
Fig.~\ref{indexdensity}. Clearly for the system sizes which we have
been able to study the free-energy per spin of most algorithms is of
order $-0.69$ (see Table~\ref{NMFalgorithms}). This is much higher than 
the value of $f$
at which $k(f)$ appears to become non-zero. But the number $-0.69$
does seem to fit better with the free energy at which index 1 saddles
begin to outnumber the index 0 saddles i.e. the mimima, so perhaps a
modified version of the `saddles rule' hypothesis does apply to
the NMF landscape.

As   mentioned   above,  the   TAP   free-energy  landscape   contains
exponentially  large (in  $N$)  numbers of  minima  and $K=1$  saddles
\emph{only}. Not  only that, but  these turning points come  in pairs,
i.e., a  minimum that is  close to  a saddle in  the sense that  it is
separated   by  a   total   free  energy   difference  that   vanishes
approximately    as   $N^{-0.26}$    in    the   limit    $N\to\infty$
\cite{CGP04}.  What this  means  in  particular is  that  for the  TAP
equations the complexity of  minima $\Sigma_0(f)$ and of $K=1$ saddles
$\Sigma_1(f)$  are \emph{equal}  for all  free energies  $f$  (and all
other  complexities  are  zero).   Therefore there  is  no  `crossing
point' at which minima begin to outnumber saddles as there is for the
NMF  equations.  Since  the  solutions we  found  are compatible  with
tracking down  states of lower free  energy, it would  appear that the
fact that there  are equal numbers of saddles and  minima over a range
of free  energy means that  one does not  get trapped in  a particular
minimum,  but  somehow manages  to  explore  further  the free  energy
landscape  by  passing  over  the nearby  saddle-point.   This  latter
statement  is,  however,  as  yet somewhat  speculative  and  requires
further investigation.

\section{Dynamical properties of the TAP iteration}
\label{sec:TAP} 

As mentioned previously, the TAP equations are usually very difficult
to solve numerically.  However, using the modified map
Eq.~\eqref{bettermap} with $\alpha>1$ surprisingly leads to a vastly
increased probability of finding solutions.  In this section we
explain in more detail why this is the case and how the \textit{ad
hoc} parameter $\alpha$ should be chosen for optimal results. First,
however, we discuss the details of the implementation as there are a
number of potential pitfalls.

\subsection{Numerical iteration}

The  iteration  Eq.~\eqref{bettermap} is  best  applied  to each  spin
separately and in sequence to improve convergence. At low temperatures
the converged  result usually contains  some $m_i$ which are  equal to
$\pm  1$ within  numerical accuracy.  While this  might  be considered
merely an irrelevant numerical inaccuracy,  it is a serious problem as
soon as the Hessian,  Eq.~\eqref{hessian}, of the solution is required
as it  is clearly  ill-defined if $m_i=\pm  1$. Therefore  we continue
iteration  of a  transformed  version  of the  TAP  equations for  the
variables $x_i =-\mathrm{sign}(m_i)\ln(1-m_i^2)$,
\begin{align}
x_i^{(k+1)}     &=     x_i^{(k)}+     \alpha\left[2\mathcal{H}_i     +
2\mathrm{sign}(\mathcal{H}_i)
\ln\frac{1+e^{-2|\mathcal{H}_i|}}{2}-x_i^{(k)}\right],
\end{align}
where  $\mathcal{H}_i =  \tanh^{-1}\G_i$.  This  particular  choice of
transformed  variables  is  adapted   to  the  form  of  the  Hessian,
Eq.~\eqref{hessian}. The diagonal elements of $H_{ij}$ can be obtained
from  $|x_i|$  simply  by  exponentiation without  fear  of  numerical
overflow or loss of accuracy.  The initial condition for the iteration
of the  transformed variables  is obtained from  the set  of converged
$m_i$  via  $x_i^{(0)}=2\mathcal{H}_i +  2\mathrm{sign}(\mathcal{H}_i)
\ln\frac{1+e^{-2|\mathcal{H}_i|}}{2}$   applied   to   all  spins   in
parallel.  This  is  important  since application  in  sequence  would
implicitly carry out  more iteration steps and would  not result in an
accurate  translation  of the  $m_i$  into  the transformed  variables
$x_i$,  thus in  practice  often leading  to  exit from  the basin  of
attraction.

When  convergence  is reached  using  the  transformed variables,  the
Hessian is calculated. Diagonalization  of it is, however, hampered by
its severe ill-conditionedness,  particularly at low temperatures. The
eigenvalues  may vary  over  100  orders of  magnitude,  owing to  the
extremely large variation of  the diagonal entries. Therefore standard
diagonalization  packages  fail  and  one  has  to  resort  to,  e.g.,
diagonalization  routines  from \cite{numrec}  modified  for use  with
arbitrary precision packages such as \cite{arprec}.

\subsection{Dynamical critical point}

The fact that iterations with $\alpha>1$ generally find many solutions
of the TAP equations and iterations with $\alpha\le 1$ basically never
find   solutions  (other   than   the  trivial   one)  suggests   that
$\alpha=\alpha_c=1$   is   a   critical   point   of   the   iterative
dynamics. Just as the iterative  behavior of e.g.\ the logistic map is
changed from convergence to a stable fixed point to eventually chaotic
dynamics  via its  control parameter,  we expect  that we  have stable
fixed points for $\alpha>1$ and chaotic behavior for $\alpha\le 1$. We
test  this hypothesis  by  various means.   First  we analyze  scaling
properties which  we expect to hold  near the critical  point. We then
examine  the lengths  of limit  cycles  above and  below the  critical
point.

\subsection{Scaling properties near the critical point}

\begin{figure}
\includegraphics[width=\linewidth]{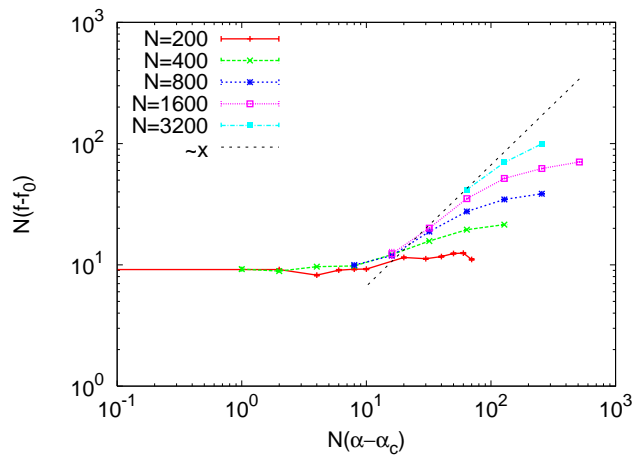}
\caption{\label{freescale}Finite-size scaling of  the free energy as a
function of  the scaling variable  $x=N(\alpha-\alpha_c)^\gamma$, with
$\gamma=1$, at inverse temperature $\beta=20$.}
\end{figure}

The choice of  the parameter $\alpha$ is not  arbitrary. A given value
of $\alpha$ favors finding a particular subset of all solutions, i.e.\
the typical  free energies found are correlated  with $\alpha$.  Under
the   hypothesis   of  the   existence   of   a   critical  point   at
$\alpha=\alpha_c=1$  the  free  energies  $f$ obtained  with  a  given
$\alpha$ are hypothesized to obey a finite-size scaling form
\begin{align}
f-f_0 &= N^{-1}F(N(\alpha-\alpha_c)^\gamma),
\end{align}
where $f_0$  is the  exact free energy  per spin in  the thermodynamic
limit,   which   is   known   exactly  from   methods   described   in
\cite{CR02,CLPR03}  to be  $f_0 \approx  -0.76324\dots$  at $\beta=20$
\cite{RizzoPrivate}.    Fig.~\ref{freescale}   shows  the   connection
between $\alpha$, the system size  $N$ and the average free energy $f$
of  the  solutions  found  at  inverse  temperature  $\beta=20$  in  a
finite-size      scaling      plot      with     scaling      variable
$x=N(\alpha-\alpha_c)^\gamma$.  The best  data  collapse is  obtained for  an
exponent $\gamma\approx  1$, but with  too much scatter to  provide an
error bar. The finite-size corrections are still very large, as can be
seen  from the deviations  at large  arguments.  The  scaling function
$F(x)$ seems  to behave  as $x^\eta$ for  large $x$  with $\eta\approx
1$. This  would imply that, away  from the critical  point, the excess
free   energy  per  spin   approaches  an   algorithm-dependent  (here
$\alpha$-dependent) value,  in accordance with a result  of Newman and
Stein \cite{newmanstein} which states that,
in the  thermodynamic limit, any  given algorithm for solving  the TAP
equations  will  find solutions  of  a  given  free energy  per  spin,
characteristic  of the  algorithm.  In  our case  any fixed  choice of
$\alpha$ defines an algorithm and  the scaling form with $\eta=1$ then
gives  $f=f_0+\text{const.}(\alpha-\alpha_c)$   for  $N\to\infty$,  as
required.   In  the  opposite   limit,  the   data  suggest   $F(0)  =
\text{const.}$,  i.e.\  for $\alpha  \to  \alpha_c$,  the excess  free
energy \emph{per spin} vanishes.

As  a spin-off of  these results  we therefore  deduce that  using the
modified map  Eq.~\eqref{bettermap} we can, in  principle, find states
with free energy  per spin arbitrarily close to  $f_0$ by choosing the
scaling variable  $x=N(\alpha-\alpha_c)^\gamma$ as small  as possible.
The drawback is  that the probability of finding  any solutions at all
goes to zero as $\alpha \to \alpha_c$. This will be illustrated in the
following two subsections.

\subsection{Cycle lengths}
\begin{figure}
\includegraphics[width=\linewidth]{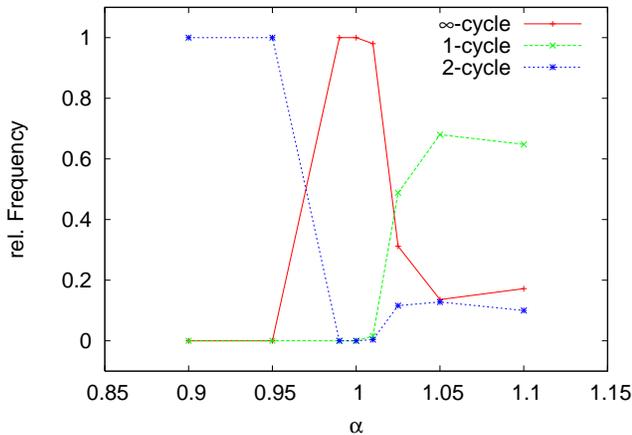}
\caption{\label{cyclefig}The data from Tab.~\ref{cycletab} plotted for
lengths 1, 2, and $\infty$}
\end{figure}

As  $\alpha$ is  decreased, limit  cycles in  the dynamics  occur with
increasing frequency.   In order to  further test the  hypothesis that
$\alpha=1$ is a critical point we have calculated the lengths of limit
cycles   as  a   function  of   $\alpha$.    Table~\ref{cycletab}  and
Fig.~\ref{cyclefig} show  the frequency with  which limit cycles  of a
given  length  were found  for  10 samples  of  an  $N=800$ system  at
$\beta=20$  using 25  different  random starting  positions each.   We
checked  for limit  cycles  up to  length  20 beyond  which they  were
considered  to be  infinitely long.   There  is a  marked change  from
$\alpha=1.1$, where 65\% of the starting positions lead to convergence
(i.e.\ a limit  cycle of length 1), to $\alpha=1$,  where 100\% of the
starting positions did not run into a limit cycle (of length less than
20) at   all.   There   is  another   very  distinct   change  between
$\alpha=0.99$ and  $\alpha=0.95$ from infinite limit  cycles to cycles
of length 2.  Closer inspection  revealed that these latter cycles are
of the form $m_i\to-m_i\to m_i$.

\begin{table*}
\caption{\label{cycletab}Frequency of appearance  of limit cycles of a
given length.  The system size is $N=800$  at temperature $\beta=20$.}
\small
\begin{tabular}{c|cccccccccc}
         & \multicolumn{10}{c}{Length of limit  cycle} \\ $\alpha$ & 1
& 2 & 3 & 4 & 5 & 6 & 7 & 8 & 9 & $\infty$ \\ \hline\hline 1.1 & 0.648
& 0.1 & 0,048 & 0,02 & 0.004 &  0 & 0 & 0.004 & 0.004 & 0.172\\ 1.05 &
0.68 &  0.128 & 0.012 & 0.028  & 0.004 & 0  & 0.004 & 0.004  & 0.004 &
0.136 \\ 1.25 & 0.312 & 0.116 & 0.012 & 0.04 & 0.008 & 0.016 & 0.004 &
0.004 & 0 & 0.488 \\ 1.01 & 0.016 & 0.004 &  0 & 0 & 0 & 0 & 0 & 0 & 0
& 0.98 \\ 1.0 & 0 & 0 & 0 & 0 & 0  & 0 & 0 & 0 & 0 & 1 \\ 0.99 & 0 & 0
& 0 & 0 & 0 & 0 & 0 & 0 & 0 & 1  \\ 0.95 & 0 & 1 & 0 & 0 & 0 & 0 & 0 &
0 & 0 & 0\\ 0.9 & 0 & 1 & 0 & 0 & 0& 0 & 0 & 0 & 0 & 0
\end{tabular}
\end{table*}

The picture  which emerges  from this data  is that for  $\alpha>1$ we
find a distribution of cycle  lengths which includes a finite fraction
of cycles  with length 1,  i.e.\ convergence.  When $\alpha$  is tuned
towards 1, more and more weight  is taken from the finite limit cycles
and transferred  to the infinite  one, until at $\alpha=1$  all cycles
have infinite  length. At the  same time, decreasing  $\alpha$ towards
the  value one  leads to  states with  ever lower  free  energies. The
optimum  results for  the  free energy  (giving  the equilibrium  free
energy per spin  in the thermodynamic limit) seemingly  occur when the
system is `at the edge of  chaos'. Similar ideas have put forward in
a  recent  paper  by  Boettcher  and Frank  \cite{BF},  who  refer  to
`optimizing at the ergodic edge.'

\subsection{Success rates}

\begin{figure}
\includegraphics[width=\linewidth]{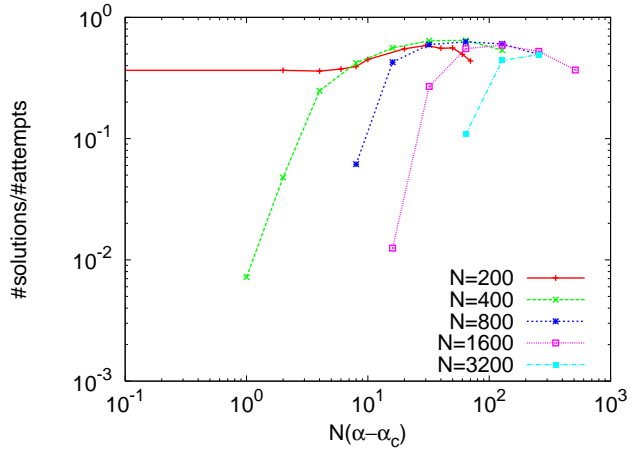}
\caption{\label{success}Success   rates  (probability  of   finding  a
solution) as a function of the scaling variable $N(\alpha-\alpha_c)$.}
\end{figure}

\begin{figure}
\includegraphics[width=\linewidth]{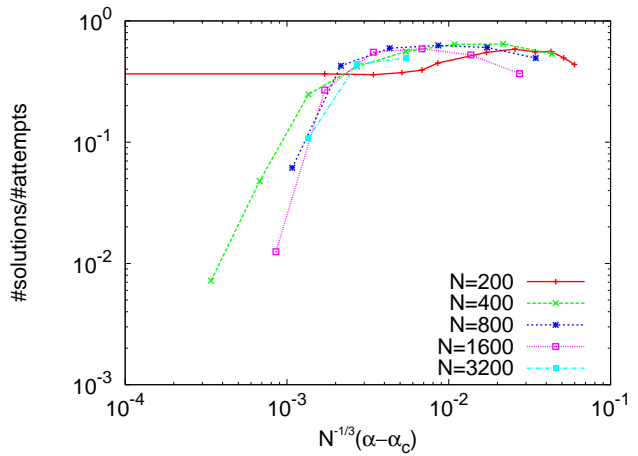}
\caption{\label{successscl}Success    rates   as    a    function   of
$N^{-1/3}(\alpha-\alpha_c)$.}
\end{figure}

The probability of finding cycles of length $1$, i.e.\ convergence, is
not only a function of $\alpha$ but, naturally, also of the system
size $N$.  While the scaling of the free energy in
Fig.~\ref{freescale} suggests that by tuning $\alpha$ towards $1$
solutions of arbitrarily low free energy can in principle be found, we
will now show that in practice this becomes increasingly difficult for
large system sizes.  Fig.~\ref{success} shows the success rates
(probability of convergence to a solution from a random starting
position) for various system sizes. The larger the system is, the more
attempts are needed to find a solution.  This statement can be made
more precise by plotting the success rates against
$x'=N^{-1/3}(\alpha-\alpha_c)$, which is done in
Fig.~\ref{successscl}. It shows that, at least for system sizes $\ge
800$, the success rates fall on a master curve which goes to zero very
quickly (faster than a power, and consistent with a Lifshitz-like
behavior $\sim e^{-\text{const.}\times {x'}^{-\zeta}}$ with
$\zeta\approx 2$) for small arguments.

So  while the appropriate  scaling variable  is $x=N(\alpha-\alpha_c)$
for  the  free  energies  and  barrier  heights  (see  below),  it  is
$x'=N^{-4/3}x$ for the success rates.  Fixing $x$ and thus a preferred
range of free energies drastically reduces the success rates for large
systems.  Assuming a  Lifshitz tail  as suggested  above,  the success
rates are proportional to $e^{-\text{const.}\times N^{4\zeta/3}}$.

\subsection{Barrier heights}

\begin{figure}
\includegraphics[width=\linewidth]{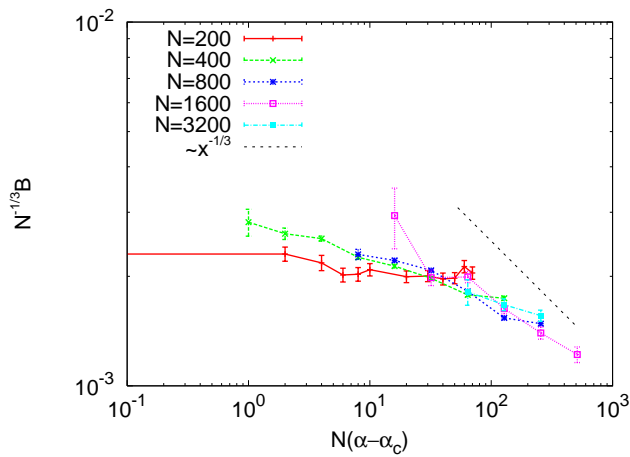}
\caption{\label{barrierscale}Finite-size scaling of the barrier height
as  a  function  of  the scaling  variable  $x=N(\alpha-\alpha_c)$  at
inverse temperature $\beta=20$.}
\end{figure}

\begin{figure}
\includegraphics[width=\linewidth]{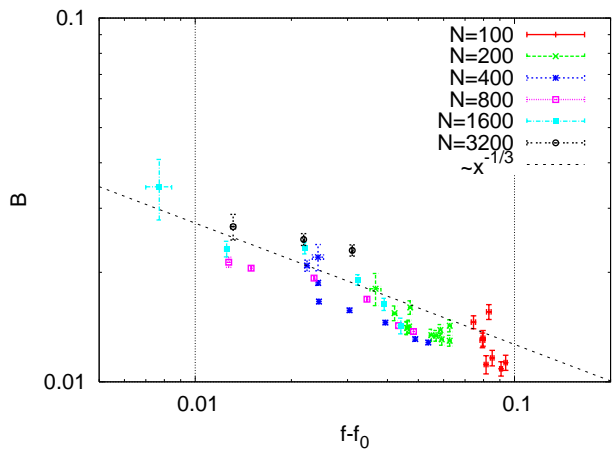}
\caption{\label{barrierfree}The (total)  barrier height as  a function
of free energy (per spin). Each data point represents the average over
all solutions found with a particular value of $\alpha$.}
\end{figure}

Since minima and  saddles of index 1 always occur in  pairs in the TAP
landscape  \cite{ABM04},  the barrier  height  of  a  solution can  be
defined  as the  free  energy  difference between  a  minimum and  its
corresponding saddle.  Fig.~\ref{barrierscale} shows a scaling plot of
the  barrier heights  using  the  same scaling  variable  as in  Fig.\
\ref{freescale}.   The  data  collapse   is  not  very  good,  but  is
consistent  with a barrier  height exponent  $\psi=1/3$, which  is the
expected scaling for the  SK model \cite{barrier}.  This suggests that
the  barriers between  very low-lying  mimima and  their corresponding
saddles, with free  energies per spin equal to  the equilibrium value,
scale as $N^{1/3}$ for large  $N$, but with a rather small coefficient
of order $10^{-3}$.  For larger  values of the scaling variable, where
the free energy  per spin lies above the  equilibrium value (see Fig.\
\ref{freescale}), the  barriers decrease.  The straight line  in Fig.\
\ref{barrierscale} has a  slope $-1/3$, so if the  data were to follow
this line  it would indicate that the  barriers become $N$-independent
at higher free  energies. Recall that the barriers  averaged over {\em
all} TAP states  were found to actually decrease  with $N$, roughly as
$N^{-0.26}$  \cite{CGP04}.   It should  be  noted,  however, that  the
states  of a  given free  energy reached  by our  particular algorithm
almost certainly  do not sample {\em  all} states of  that free energy
uniformly.

The  scaling behavior  of the excess free energy  per spin and of the 
barrier heights  for large argument  $x=N(\alpha-\alpha_c)$, namely  
$f - f_0 \sim x$ and $N^{-1/3}B\sim x^{-1/3}$, indicates  that (at least 
for large $x$) the  barrier height  is directly  proportional to  
$(f-f_0)^{-1/3}$  with no dependence   on  $N$   or   $\alpha-\alpha_c$.  
Fig.~\ref{barrierfree} demonstrates that surprisingly this is not only 
true for large $x$ but for \textit{all} values of $N$ and 
$\alpha-\alpha_c$ explored. This is an important  result since the  
dependence on the  unphysical $\alpha$ has been eliminated and the 
relation $B\sim (f-f_0)^{-1/3}$ is a property of physical quantities 
alone.

\subsection{Eigenvalues of the Hessian}

\begin{figure}
\includegraphics[width=\linewidth]{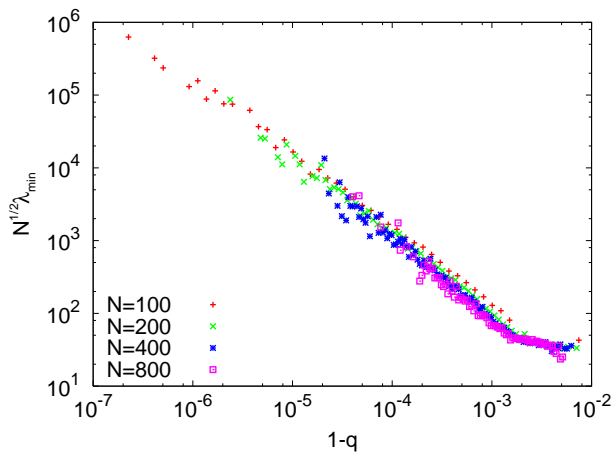}
\caption{\label{minevalq}Finite  size  scaling  plot  of  the  average
smallest   eigenvalue   $\overline{\lambda_{\text{min}}(q)}$  of   the
Hessian  at  a  minimum  versus  $1-q$.  The  eigenvalues  scale  with
$N^{-1/2}$.  The data  shown is  for  $\alpha$ such  that the  scaling
variable is a constant, namely $N(\alpha-\alpha_c)=16$.}
\end{figure}

According  to   our  picture  of   the  TAP  landscape   described  in
\cite{ABM04}, the  Hessian at a  minimum should, in  the thermodynamic
limit, have one null eigenvalue  and a band of eigenvalues starting at
a value strictly  larger than zero. In order to  confirm this, we have
studied the eigenvalues  of the Hessian at the  solutions found by our
algorithm. Since the arbitrary-precision diagonalization is forbidding
for large system sizes, we are  restricted to $N\le 800$. In the range
of accessible  system sizes we find  a huge variation  in the smallest
eigenvalue  $\lambda_{\text{min}}$ (see Fig.~\ref{minevalq})  which is
at  first  sight contradicting  our  landscape  picture. However,  the
eigenvalues $\lambda_{\text{min}}$ show  a strong correlation with the
corresponding value of $q$ at the minimum. This is understandable from
the  form  of  the  Hessian, Eq.~\eqref{hessian}  since  the  smallest
eigenvalue  will be of  the same  order of  magnitude as  the smallest
diagonal    entry   of    $H_{ij}$,   which    is    proportional   to
$1/(1-m_{\text{min}}^2)$  ($m_{\text{min}}$ being the  $m_i^*$ closest
to 0).  Therefore, for $q$ very close  to 1, $\lambda_{\text{min}}\sim
1/(1-m_{\text{min}}^2)\sim 1/(1-q)$.  Fluctuations of $q$  towards $1$
will therefore cause  great fluctuations in $\lambda_{\text{min}}$. In
the thermodynamic  limit, $q$ should be  self-averaging, therefore the
variations in $\lambda_{\text{min}}$  become small and our observation
merely indicates the enormous finite-size corrections still present at
our accessible system sizes.

The correlation between $\lambda_{\text{min}}$  and $q$ and the strong
variation of  $\lambda_{\text{min}}$ over several  orders of magnitude
indicate that a simple  average of $\lambda_{\text{min}}$ might not be
very    informative.      Instead    we    consider     the    average
$\overline{\lambda_{\text{min}}(q)}$ of the  eigenvalues as a function
of $q$.  This  quantity is plotted in Fig.~\ref{minevalq}  in a finite
size scaling  plot which shows  good data collapse, in  particular for
larger $N$. There are several  points to notice in this figure. First,
the eigenvalues scale with $N^{-1/2}$  but can still be very large (of
the order $1000$ or more at  the very low temperature, $\beta =20$, at
which we are working).  Second, for larger $N$ the range of $q$-values
shrinks. This  must be  the case as  in the thermodynamic  limit there
should   be   only  one   value   of   $q$.   Third,  the   range   of
$\lambda_{\text{min}}$            also           shrinks           and
$\overline{\lambda_{\text{min}}}$ (where the  average now includes all
values of  $q$) goes to  zero as $N^{-1/2}$  (note that it  would have
been   very  difficult   to   obtain  this   result   from  a   simple
$q$-independent  average  of   $\lambda_{\text{min}}$).   This  is  in
support  of the  picture of  the  TAP free  energy landscape  outlined
previously \cite{ABM04} which asserts that minima and saddles occur in
pairs  which  move closer  together  as  the  system size  grows.  The
smallest eigenvalues of the Hessian  therefore have to go to zero with
increasing $N$.

\begin{figure}
\includegraphics[width=\linewidth]{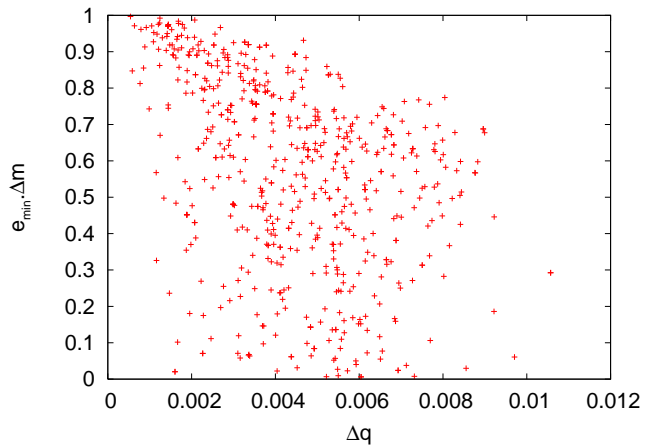}
\caption{\label{overlap}Overlap      between      the      eigenvector
$e_{\text{min}}$ of  the Hessian belonging to  the smallest eigenvalue
and the normalized difference vector  $\Delta m$ between a minimum and
the    corresponding    saddle    as    a    function    of    $\Delta
q=|q_{\text{saddle}}-q_{\text{min}}|$.  The data  are  for $N=800$  at
$\beta=20$ and $\alpha=1.02$.}
\end{figure}

In order to provide further  support for this picture, we have plotted
in   Fig.~\ref{overlap}   the    overlap   between   the   eigenvector
$e_{\text{min}}$  belonging   to  the  smallest   eigenvalue  and  the
normalized difference  vector (in  $m$-space) between the  minimum and
its  corresponding saddle $\Delta  m$, as  function of  the difference
$\Delta  q=|q_{\rm  saddle}-q_{\rm min}|$.  If  the  two vectors  were
uncorrelated,  the distribution  of  their overlaps  would be  sharply
peaked around zero. The figure  shows, however, that they are strongly
correlated  and  that  the   direction  of  the  smallest  eigenvector
coincides  with the  direction in  which the  corresponding  saddle is
found.  The  correlation is particularly strong  for small differences
of $q$  between saddle  and minimum since  these pairs are  very close
together.

\begin{figure}
	\includegraphics[width=\linewidth]{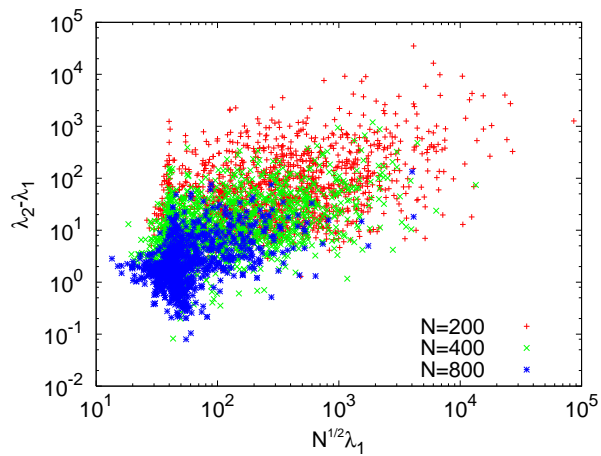}
    \caption{\label{lambda12}Plot of the difference between the second
 and     the     lowest     eigenvalue     as    a     function     of
 $N^{1/2}\lambda_{\text{min}}$. }
\end{figure}

The second smallest  eigenvalue $\lambda_2$ is expected to  be, in the
thermodynamic limit, strictly larger than  $0$. In order to test this,
we  have   plotted  in  Fig.~\ref{lambda12}   the  difference  between
$\lambda_2$    and   $\lambda_{\text{min}}$    as   a    function   of
$N^{1/2}\lambda_{\text{min}}$. In the thermodynamic limit there should
be only a horizontal line of finite width in this plot. Clearly we are
still very far  away from this limit, even  for $N=800$, the variation
in   $\lambda_2-\lambda_{\text{min}}$   being   of   order   $10$   or
more. However,  it is also clear  that the data becomes  more and more
concentrated  for  larger  $N$.   Observe  also  the  emergence  of  a
horizontal line for $N=800$.

\begin{figure}
\includegraphics[width=\linewidth]{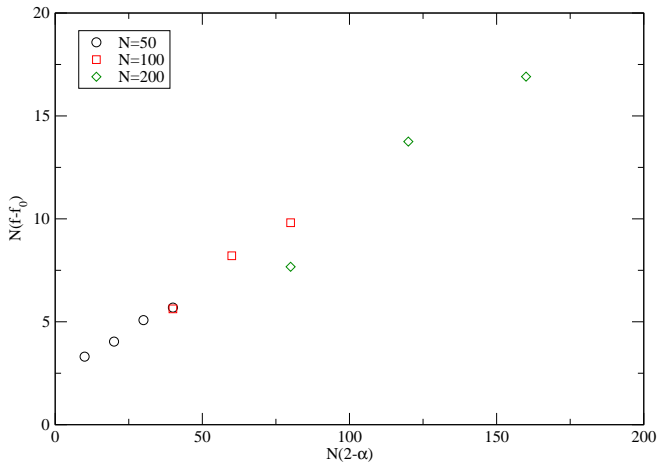}
\caption{\label{ZeroTEnergiesShifted}   Finite-size  scaling   of  the
NMF free-energy  at   $T=0$  as  a   function  of  the   scaling  variable
$x=N(2-\alpha)$.  }
\end{figure}

\section{Discussion and Summary}
\label{discussion}
In  this  work  we   have  compared  and  contrasted  the  free-energy
landscapes  of  two  closely  related  models, described  by  the  TAP
equations  and the  NMF  equations respectively.  The TAP  free-energy
surface has a  simple structure with turning points  of only two types
---minima, and saddle  points of index 1.  Furthermore,  the minima and
saddle-points occur in  pairs, each containing a minimum  and a saddle
point  that are  close neighbors  in configuration  space. By  using a
novel algorithm controlled by a single parameter $\alpha$, we are able
to access TAP  states whose free energy per spin is  equal to the {\em
equilibrium free  energy per spin}  at the given temperature,  with an
excess {\em total} free energy of order 10 (see Fig.~\ref{freescale}).
This  is  achieved by  choosing  the  parameter  $\alpha$ close  to  a
critical  value,  $\alpha_c$,  beyond  which  the  iterative  dynamics
becomes  chaotic.  The  dependence  of the  excess  total free  energy
$\Delta F$  on $N$ and $\alpha$  is broadly consistent  with a scaling
form  $\Delta  F =  F[N(\alpha-\alpha_c)^\gamma]$,  valid in the limit 
$N\to\infty$, $\alpha\to\alpha_c$ with  $N(\alpha-\alpha_c)^\gamma$ 
fixed, and $\gamma  \approx 1$.  The scaling function $F(x)$ should
behave as $F(x) \sim x$ for large $x$, so that $\Delta F$ becomes of 
order $N$  away from the critical point, while $F(0)$ is a non-zero 
constant. 

We have also studied the free-energy barriers,  given by the free-energy 
difference between a minimum and its corresponding saddle point. In the 
low free-energy regime where $N(\alpha-\alpha_c)^\gamma \ll 1$, 
the barriers seem to grow as $N^{1/3}$, albeit with a small coefficient, 
while for $N(\alpha-\alpha_c)^\gamma \gg 1$ we expect the barriers to be 
of order unity or smaller (see Fig.~\ref{barrierscale}). This leads to 
a picture where only the lowest free energy states (with free energies 
per spin equal to the equilibrium one) are separated by barriers which 
increase  in some manner with $N$.
 
In contrast to the TAP free-energy function, the NMF free energy has 
saddle points with index $K$ taking values up to order $N$. The separate 
complexities $\Sigma_K(f)$ have been computed numerically (see Fig.\ 
\ref{MultiComplexity}). The free energy at which the minima start to 
outnumber (exponentially in $N$) the saddle points seems to play the role
of a threshold free energy which acts as lower bound on the free energy 
that is dynamically accessible (Fig.~\ref{CrossingPoints}).

We conclude with with some general remarks concerning optimization
problems.  If one is faced with a combinatorial optimization problem,
such as finding the ground state of a spin glass, one approach is to
write down TAP-like equations and solve them using an iterative scheme
which is related to a descent procedure on the free-energy landscape.
As the temperature is lowered the states obtained approach those of a
zero-temperature problem.  In the spin-glass context, it does not seem
to matter, {\em a priori}, whether the finite-temperature
generalization of the problem is through the TAP equations or the NMF
equations, since they become identical at $T=0$. We believe, however,
that the very different structures of the free-energy landscapes in
the two models will make a very big difference in practice.  While the
NMF free-energy surface is `robust', with its threshold free energy
and, presumably, large free energy barriers between states, the TAP
free energy surface may be termed `fragile'.  With any such problem
therefore, it will pay to construct finite-temperature equations
corresponding to a free energy with a fragile structure.  This process
might be called `sculpting the free-energy landscape'.

Finally we would like to point out that using algorithms which take
one off the free-energy surface into physically unreachable regions
before reaching the final physical fixed point may also have great
utility, as illustrated by our `edge of chaos' algorithm.  The success
of this type of algorithm seems to be partly due to the fact that by
allowing unphysical values of the $m_i$ one can tunnel through
barriers and partly because tuning the adjustable parameter $\alpha$
towards the edge of chaos yields the solutions of low free energy
which tend to be surrounded by the highest barriers which can trap the
algorithm into a fixed point solution.  We mainly used the `edge of
chaos' algorithm in connection with solving the TAP equations but it
seems to work well also for the NMF equations.  In
Fig.~\ref{CrossingPoints} we have included the results obtained by
setting $\alpha=1.9$ which is close to the `edge of chaos' but
without attempting to tune it to its optimum value for each value of
$N$. (For the NMF equations the edge of chaos seems to be close to 2
and the values of $\alpha$ which give stable solutions are less than
the critical value).  It is clear that it outperforms the other
iterative algorithms, which approximate to descents on the free-energy
surface, by a considerable margin.  At $T=0$ the algorithm is
especially simple and Fig.~\ref{ZeroTEnergiesShifted} shows that it
seems to work as efficiently as the `edge of chaos' algorithm applied
to the TAP equations at low temperatures. However, it does suffer from
similar problems as the `edge of chaos' algorithm when applied to the
TAP equations: we noticed that the number of iterations required to
achieve convergence to a fixed point increased dramatically as the
system size $N$ increases, or as $\alpha$ approaches its critical
value.

In the context of mean-field problems such as finding the groundstate
of the SK model, the `edge of chaos' algorithm works well. We have
also tested it on the one-dimensional spin glass at $T=0$, (which can
of course be solved exactly, enabling one to easily judge the accuracy
of any proposed treatment).  Although it again outperformed simple
descent algorithms it was hard to get solutions whose
energies per spin were equal to the equilibrium energy per spin.  
We attribute the difference to the nature of the free-energy
landscape. In one dimension the barriers around the states of the
lowest energy are little different in height to those surrounding
metastable states of much higher free energy.

\begin{acknowledgments}
RAB acknowledges an EPSRC Fellowship GR/R44768 under which part of
this work was conducted, and a Royal Society of Edinburgh Personal
Research Fellowship under which it was completed.  TA acknowledges
support by the DFG under grant Zi~209/7. We thank G.\ Jentsch for
kindly performing some of the simulations used in this work.
\end{acknowledgments}

\end{document}